\UseRawInputEncoding

\documentclass[reprint, twocolumn, showkeys]{revtex4}

\usepackage{graphicx}
\usepackage{dcolumn}
\usepackage{bm}
\usepackage{color}
\usepackage{amsmath,amsthm}
\usepackage{amssymb,bm}
\usepackage{mathrsfs}
\usepackage{comment}
\usepackage{ulem}

\graphicspath{%
    {converted_graphics/}
    {/}
}
\begin{document}

\title{Floquet-engineered chiral-induced spin selectivity}
\author{Nguyen Thanh Phuc}
\email{nthanhphuc@moleng.kyoto-u.ac.jp}
\affiliation{Department of Molecular Engineering, Graduate School of Engineering, Kyoto University, Kyoto 615-8510, Japan}

\begin{abstract}
The control of electron spin, which is crucial to the stability of matter, offers new possibilities for manipulating the properties of molecules and materials with potential applications in spintronics and chemical reactions. 
Recent experiments have demonstrated that the electron transmission through chiral molecules depends on the electron spin orientation, a phenomenon known as chiral-induced spin selectivity (CISS).
In this study, we show that CISS can be observed in achiral systems driven by an external circularly polarized laser field in the framework of Floquet engineering.
By using the Floquet theory for a time-periodically driven system to investigate spin-dependent electron transport in a two-terminal setup, we demonstrate that the spin polarization can approach unity if the light intensity is sufficiently strong, the rate of dephasing is sufficiently small, and the average chemical potential of the two leads is within an appropriate range of values, which is narrow because of the high frequency of the laser field.
To obtain a wider range of energies for large spin polarization, a combination of chiral molecules and light-matter interactions is considered and the spin polarization of electrons transported through a helical molecule driven by a laser field is evaluated.
\end{abstract}


\maketitle

\section{Introduction}
\label{sec: Introduction}
The spin angular momentum is a fundamental characteristic of the electron which can regulate the properties of atoms, molecules, and materials. 
In accordance with their quantum statistics, two electrons occupying the same orbital must have opposite spins, namely the Pauli exclusive principle.
The exchange energy that determines the energy gap between spin singlet and triplet states is particularly important for understanding the electronic configurations and the accompanying magnetic and optical properties of different atoms and molecules.
The recently discovered effect of chiral-induced spin selectivity (CISS)~\cite{Ray99} offers a new method for controlling electron spin in chiral molecules and materials.
In CISS, electron spins are polarized when they are transported through chiral molecules, and the spin polarization depends on the handedness of the molecule~\cite{Naaman12, Naaman15, Naaman19}.
CISS has been extensively studied~\cite{Medina15, Diaz18, Geyer19, Michaeli19, Fransson19, Zollner20, Liu21, Kato22} and observed in a wide range of molecules and materials, including DNA, bacteriorhodopsin, helicenes, and others~\cite{Xie11, Carmeli02, Gohler11, Mishra13, Kettner15, Kettner18, Lu19, Mishra19, Jia20}.
Applications of CISS go beyond those in the field of spintronics~\cite{Dor17, Suda19} as it can be employed for enantiomer selection~\cite{Rosenberg15, Ghosh18} and promotion of chemical reactions, such as electrochemical water splitting~\cite{Mtangi15, Mtangi17}.
Using an argument based on Kramer's degeneracy, it can be shown that the spin polarization in the linear transport through a single-orbital-per-site and two-terminal chiral system would vanish unless the time-reversal symmetry is broken~\cite{Kiselev05, Bardarson08, Gutierrez13, Utsumi20}. 
Non-unitary effects, such as dephasing or leaking, can effectively break the time-reversal symmetry~\cite{Guo12, Guo14, Matityahu13, Matityahu16}.
However, the effect of dephasing tends to disrupt the quantum interference across different transport channels, which is essential to CISS as implicitly indicated by the fact that CISS cannot be detected in a single-orbital-per-site single-helix model without an electron's long-range tunneling~\cite{Gohler11, Mishra13,Guo12, Guo14}. 

It has recently been demonstrated that CISS can be observed in achiral molecules and materials strongly coupled to a circularly polarized mode of an optical cavity or waveguide~\cite{Phuc23}. 
Here, the optical degree of freedom was treated quantum mechanically, the light-matter coupling is strong at the single-photon level, and the quantum fluctuation in the vacuum state of the optical mode can give rise to CISS in the complete absence of an external light field.
In this study, we consider a conventional setup of light-matter interaction in which an external laser is applied to an electronic system. 
The effect of laser field on electrons is treated classically in terms of a time-periodic potential, that is in the framework of Floquet engineering~\cite{Bukov15, Eckardt17, Moessner17, Oka19, Weitenberg21}.
It has been proposed and demonstrated that Floquet engineering can be used to control charge and energy transports~\cite{Dakhnovskii95, Kohler05, Phuc18, Phuc19}, manipulate quantum phase transitions~\cite{Zenesini09}, generate artificial magnetic fields for charge-neutral particles~\cite{Aidelsburger11, Struck12, Aidelsburger13, Miyake13}, and create topologically nontrivial band structures~\cite{Oka09, Jotzu14, McIver19}.  
In this study, we show that the spin selectivity can be observed in electron transport through achiral molecules and materials driven by an external circularly polarized laser field.
Unlike the ordinary CISS in chiral molecules, the spin polarization under Floquet engineering can be nonzero without dephasing.
This is because the light-matter interaction can effectively break the time-reversal symmetry in the dynamics of electrons, no matter if light is treated in a quantum or classical manner.
The Floquet theory is applied to investigate spin-dependent electron transport in a two-terminal setup~\cite{Kohler05}.
The spin polarization is found to approach unity if the average chemical potential of the two leads is within an appropriate range of values, which is narrow because of the high frequency of the laser.
Chiral molecules and light-matter interactions can be combined to produce substantial spin polarization over a wider energy range. 
This is demonstrated by evaluating the spin polarization of the electrons transported through a helical molecule under the influence of a circularly polarized laser field.
How the spin polarization varies with molecular length, driving frequency and amplitude, and dephasing rate is investigated.

\section{Floquet-engineered CISS}
\label{sec: Floquet-engineered CISS}
\begin{figure}[tbp] 
  \centering
  \includegraphics[width=3in, keepaspectratio]{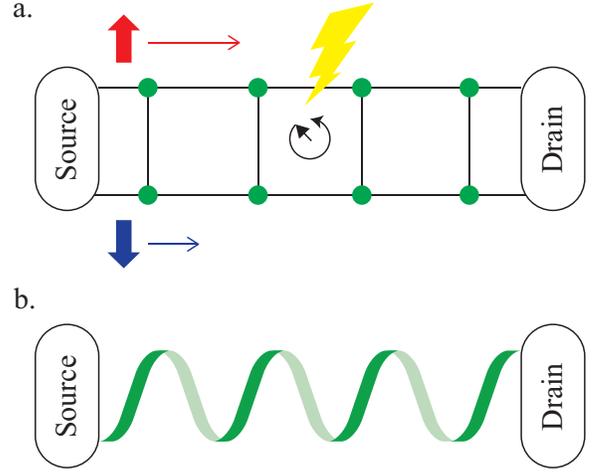}
  \caption{Illustration of the (a) Floquet-engineered CISS in achiral matters and (b) ordinary CISS in chiral molecules and materials. In panel a, electrons being transported through an achiral system, such as a square two-dimensional lattice, are time-periodically driven by a circularly polarized laser field, whose vector potential rotates in the counter-clockwise direction. The two-terminal configuration in which the system is connected to a source and a drain is considered in both panels a and b. The spin polarization is reflected by the difference in the transmission of electrons with one spin orientation over the other.}
  \label{fig: system}
\end{figure}

As an example of achiral systems, we consider spin-dependent electron transport through a two-dimensional (2D) square lattice (of size $N_x\times N_y$ in the $x-y$ plane) in a two-terminal setup, as illustrated in Fig.~\ref{fig: system}a.
With the potential gradient pointing along the $z$ direction, the Rashba spin-orbit coupling (SOC) has the following form: $\hat{H}_\text{SO}=\alpha(\hat{\sigma}_x\hat{p}_y-\hat{\sigma}_y\hat{p}_x)$, where $\alpha=-(e\hbar/4m_\text{e}^2c^2)(\text{d}V/\text{d}z)$, and $\hat{\boldsymbol{\sigma}}$ and $\hat{\mathbf{p}}$ are the spin Pauli matrices and momentum operators of electrons, respectively.
An external circularly polarized laser field with the time-dependent vector potential given by $A_x(t)=A_0\cos\Omega t$, $A_y(t)=A_0\sin\Omega t$, and $A_z(t)=0$ is applied to the system.
Here, $A_0$ is the amplitude of the vector potential. 
The spatial dependence of the vector potential can be ignored as the system is typically small compared to the optical wavelength.
The interaction between electrons and the laser field is taken into account in the lattice model by using Peierls substitution~\cite{Peierls33}, where the tunneling amplitude between two sites centered at $\mathbf{r}_i$ and $\mathbf{r}_j$ is multiplied by a factor $\exp\left\{(ie/\hbar)\int_{\mathbf{r}_i}^{\mathbf{r}_j}\mathbf{A}(t)\cdot\text{d}\mathbf{r}\right\}$.
Here, the path integral is performed along the line connecting the two sites.
The time-dependent Hamiltonian of the laser-driven electron system is then given by
\begin{align}
\hat{H}(t)=&\sum_{j_x=1}^{N_x-1} \sum_{j_y=1}^{N_y}\sum_{s,s'}\Big\{\hat{c}_{j_x+1,j_y,s}^\dagger
\left[t_x(t)\delta_{ss'}+i t_x^\text{SO}(t)(\sigma_y)_{ss'}\right] \nonumber\\
&\times \hat{c}_{j_x,j_y,s'} +\text{h.c.}\Big\} 
+\sum_{j_x=1}^{N_x} \sum_{j_y=1}^{N_y-1}\sum_{s,s'}\Big\{\hat{c}_{j_x,j_y+1,s}^\dagger \nonumber\\
&\times \left[t_y(t)\delta_{ss'}-i t_y^\text{SO}(t)(\sigma_x)_{ss'}\right]
 \hat{c}_{j_x,j_y,s'} +\text{h.c.}\Big\},
\end{align}
where 
\begin{align}
t_x^{(\text{SO})}(t)=&t_x^{(\text{SO})}e^{-i\mathcal{A}_x\cos\Omega t},\\
t_y^{(\text{SO})}(t)=&t_y^{(\text{SO})}e^{-i\mathcal{A}_y\sin\Omega t}.
\end{align}
Here, $\hat{c}_{j_x,j_y,s}$ denotes the annihilation operator of an electron with spin $s$ at site $(j_x,j_y)$;
$t_{x,y}$ and $t_{x,y}^\text{SO}$ represent the spin-conserved and SOC-induced tunneling amplitudes, respectively, which are proportional to $\langle\psi_{j_x+1,j_y}|\nabla_x|\psi_{j_x,j_y}\rangle$ and $\langle\psi_{j_x,j_y+1}|\nabla_y|\psi_{j_x,j_y}\rangle$ [$\psi_{j_x,j_y}$ is the Wannier wave function at site $(j_x,j_y)$]; 
$\mathcal{A}_{x,y}=eA_0d_{x,y}/\hbar=eE_0d_{x,y}/\hbar\Omega$ are the dimensionless driving amplitudes in the $x$ and $y$ directions ($E_0$ is the electric field amplitude, and $d_x$ and $d_y$ are the lattice constants); 
h.c. refers to Hermitian conjugate; without loss of generality, the site energy is set to zero.

Using the Floquet theory, the average current of electrons with spin $s=\uparrow, \downarrow$ in the two-terminal setup can be expressed as~\cite{Kohler05} 
\begin{align}
I_s=&\frac{e}{h}\sum_{k=-\infty}^\infty \int \text{d}\epsilon \sum_{s'=\uparrow,\downarrow}
\Big[ T_{\text{R}s, \text{L}s'}^{(k)}(\epsilon) f_{\text{L}}(\epsilon) \nonumber\\
-& T_{\text{L}s, \text{R}s'}^{(k)}(\epsilon) f_{\text{R}}(\epsilon)\Big],
\end{align}
where $f_\text{L,R}(\epsilon)$ are the Fermi-Dirac distribution functions for electrons in the left and right leads, respectively. 
The transmission probability from the left to the right lead for electrons with initial spin $s'$, initial energy $\epsilon$ and final spin $s$, final energy $\epsilon+k\hbar\Omega$, that is, the probability for a scattering event under the absorption (emission) of $|k|$ photons if $k>0$ ($k<0$), is given in terms of the broadening functions $\Gamma_\text{L,R}$ and the Fourier components $G^{(k)}$ of the retarded Green's function by
\begin{align}
T_{\text{R}s, \text{L}s'}^{(k)}(\epsilon)=&\Gamma_{\text{R}}(\epsilon+k\hbar\Omega) \Gamma_{\text{L}}(\epsilon)\nonumber\\
&\times \left| \sum_{j_y,j_y'=1}^{N_y}G_{j_x=1, j_y, s'; j_x'=N, j_y', s}^{(k)}(\epsilon)\right|^2.
\end{align}
Similarly, the transmission probability for electrons from the right to the left lead is given by
\begin{align}
T_{\text{L}s, \text{R}s'}^{(k)}(\epsilon)=&\Gamma_{\text{L}}(\epsilon+k\hbar\Omega) \Gamma_{\text{R}}(\epsilon)\nonumber\\
&\times \left| \sum_{j_y,j_y'=1}^{N_y}G_{j_x=N, j_y, s'; j_x'=1, j_y', s}^{(k)}(\epsilon)\right|^2.
\end{align}
The wide-band limit is taken by assuming that the broadening functions are energy independent: $\Gamma_\text{R,L}(\epsilon)\to \Gamma_\text{R,L}$. 
If  a bias voltage $V_\text{b}$ is applied, the chemical potentials of the two leads are given by $\mu_\text{L,R}=\bar{\mu}\pm eV_\text{b}/2$ where $\bar{\mu}$ is the average chemical potential.
At zero temperature, the Fermi-Dirac distribution function reduces to the step function, and the differential conductance $c_s\equiv(\text{d}I_s/\text{d}V_\text{b})_{V_\text{b}=0}$ is given by
\begin{align}
c_s=\frac{e^2}{2h}\sum_{k=-\infty}^\infty \sum_{s'=\uparrow,\downarrow}
\left[T_{\text{L}s, \text{R}s'}^{(k)}(\bar{\mu})+T_{\text{R}s, \text{L}s'}^{(k)}(\bar{\mu})\right].
\end{align}
The spin polarization is defined by 
\begin{align}
P_\text{s}=\frac{c_\uparrow-c_\downarrow}{c_\uparrow+c_\downarrow}.
\end{align}

The Fourier components of the Green's function can be expressed in terms of the Fourier components of the Floquet states as
\begin{align}
G^{(k)}(\epsilon)=\sum_{\alpha,k'}
\frac{|u_{\alpha,k'+k}\rangle\langle u_{\alpha,k'}^+|}
{\epsilon-(\epsilon_\alpha+k'\hbar\Omega-i\hbar\gamma_\alpha)},
\end{align}
where the Fourier components $|u_{\alpha,k}\rangle$ of the Floquet state $|u_\alpha(t)\rangle$ are defined by
\begin{align}
|u_\alpha(t)\rangle=&\sum_k |u_{\alpha,k}\rangle e^{-ik\Omega t},\\
|\psi_\alpha(t)\rangle=&\,e^{(-i\epsilon_\alpha/\hbar-\gamma_\alpha)t}|u_\alpha(t)\rangle.
\end{align}
Here, the state $|\psi_\alpha(t)\rangle$ satisfies the time-dependent Schrodinger equation, from which the Floquet state $|u_\alpha(t)\rangle$ is a solution of the eigenvalue equation in the extended Hilbert space with a periodic time coordinate
\begin{align}
\left[ \hat{H}(t)-i\hat{\Sigma}-i\hbar\frac{\text{d}}{\text{d}t}\right]|u_\alpha(t)\rangle=(\epsilon_\alpha-i\hbar\gamma_\alpha)|u_\alpha(t)\rangle.
\label{eq: Floquet eigenvalue equation}
\end{align}
The self-energy $\hat{\Sigma}$, which results from the coupling of electrons to the leads, has the following matrix elements:
\begin{align}
\Sigma_{j_x, j_y, s; j_x', j_y', s'}=\delta_{ss'}\left(\frac{\Gamma_\text{L}}{2}\delta_{j_x,1}\delta_{j_x',1}+\frac{\Gamma_\text{R}}{2}\delta_{j_x,N}\delta_{j_x',N}\right).
\end{align}
Due to the Brillouin zone structure of the Floquet spectrum, it is sufficient to compute all eigenvalues in the first Brillouin zone $-\hbar\Omega/2\leq\epsilon_\alpha <\hbar\Omega/2$. 
Since the operator on the left-hand side of Eq.~\eqref{eq: Floquet eigenvalue equation} is non-Hermitian, the eigenvalues $\epsilon_\alpha-i\hbar\gamma_\alpha$ are generally complex valued and the right eigenvectors are not mutually orthogonal. 
To determine the Green's function, it is needed to solve the adjoint Floquet eigenvalue equation
\begin{align}
\left[ \hat{H}(t)+i\hat{\Sigma}-i\hbar\frac{\text{d}}{\text{d}t}\right]|u_\alpha^+(t)\rangle=(\epsilon_\alpha+i\hbar\gamma_\alpha)|u_\alpha^+(t)\rangle.
\label{eq: adjoint Floquet eigenvalue equation}
\end{align}
The Floquet states $|u_\alpha(t)\rangle$ together with the adjoint states $|u_\alpha^+(t)\rangle$ form at equal times a complete bi-orthogonal basis:
\begin{align}
\langle u_\alpha^+(t)|u_\beta(t)\rangle=&\,\delta_{\alpha\beta},\\
\sum_\alpha |u_\alpha(t)\rangle\langle u_\alpha^+(t)|=&\,1.
\end{align}
The Floquet eigenvalue equations~\eqref{eq: Floquet eigenvalue equation} and \eqref{eq: adjoint Floquet eigenvalue equation} can be rewritten in terms of the Fourier components as
\begin{align}
\sum_j\left[ \hat{H}_{k-j}-\delta_{jk}(j\Omega\hat{I}+i\hat{\Sigma})\right]|u_{\alpha,j}\rangle=&(\epsilon_\alpha-i\hbar\gamma_\alpha)|u_{\alpha,j}\rangle,\\
\sum_j\left[ \hat{H}_{k-j}-\delta_{jk}(j\Omega\hat{I}-i\hat{\Sigma})\right]|u_{\alpha,j}^+\rangle=&(\epsilon_\alpha+i\hbar\gamma_\alpha)|u_{\alpha,j}^+\rangle,
\end{align}
where $\hat{I}$ is the identity operator in the original Hilbert space, and 
\begin{align}
\hat{H}_m=\frac{1}{T}\int_0^T \text{d}t\, e^{im\Omega t}\hat{H}(t).
\end{align}
In the following calculations, the system parameters are set to $t_x=t_y=0.1\,\text{eV}$, $t_x^\text{SO}=t_y^\text{SO}=0.012\,\text{eV}$, and $\Gamma_\text{L}=\Gamma_\text{R}=1\,\text{eV}$, which are typical orders of magnitude for electron transport in organic molecules, such as DNA and proteins~\cite{Yan02, Endres04, Senthilkumar05, Hawke10}.

\begin{figure}[tbp] 
  \centering
  \includegraphics[width=3.4in, keepaspectratio]{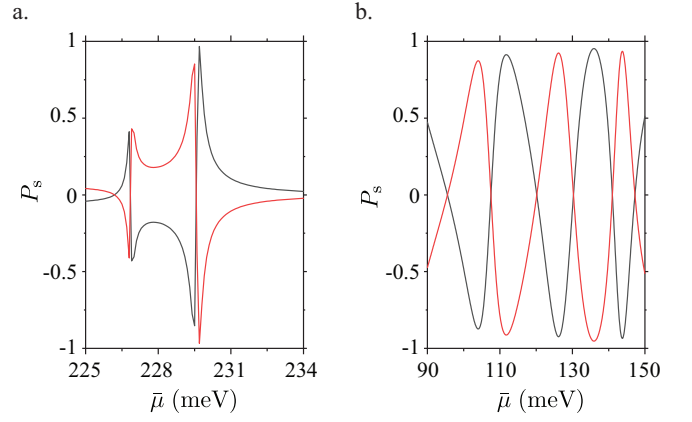}
  \caption{Spin polarization $P_\text{s}$ of electrons transported through (a) a square lattice and (b) a helical molecule in a two-terminal setup as a function of the average chemical potential $\bar{\mu}$ of the two leads. In both cases, an external circularly polarized laser field with frequency $\Omega=1\,\text{eV}/\hbar$ and dimensionless amplitude $\mathcal{A}=1$ is applied (see the text for details). The red and black curves correspond to the circular polarization in the clockwise and anticlockwise directions, respectively. Here, the chemical potential is measured relative to the site energy of electrons.}
  \label{fig: spin polarization as a function of energy}
\end{figure}

The spin polarization is shown in Fig.~\ref{fig: spin polarization as a function of energy}a as a function of the average chemical potential in the range of $225\,\text{meV}<\bar{\mu}<234\,\text{meV}$ for $N_x=30$ and $N_y=2$. 
The driving frequency and dimensionless amplitudes were set to $\Omega=1\,\text{eV}/\hbar$ and $\mathcal{A}_x=\mathcal{A}_y=1$, respectively, and the dephasing was ignored.  
It is evident that the spin polarization is non-zero without dephasing, in contrast to the ordinary CISS in chiral molecules and materials.  
This is due to the fact that the light-matter interaction effectively breaks the time-reversal symmetry in the dynamics of electrons. 
This can be understood by considering the high-frequency limit of the Floquet theory, where $\Omega$ is larger than the other system parameters. 
In this limit, the effective Hamiltonian for electrons can be derived using Van Vleck degenerate perturbation theory~\cite{Shavitt80, Eckardt15}.
Up to the first order in $1/\Omega$, the effective Hamiltonian is given by
\begin{align}
\hat{H}_\text{eff}=\hat{H}_0+\sum_{m\not=0}\frac{[\hat{H}_{-m},\hat{H}_m]}{2m\hbar\Omega}.
\end{align}
The spin-dependent effective tunneling amplitude between the $i$th and $j$th sites then reads
\begin{align}
t_{i,s;j,s'}^\text{eff}=t_{i,s;j,s'}^0+\sum_{m\not=0}\sum_{k\not=i,j}\sum_{s''}
\frac{t_{i,s;k,s''}^{-m}t_{k,s'',j,s'}^m}{m\hbar\Omega}.
\label{eq: spin-dependent effective tunneling amplitude}
\end{align}
The Fourier components of the effective tunneling amplitudes in the $x$ and $y$ directions are given by
\begin{align}
t_x^{(\text{SO})m}=&\frac{1}{T}\int_0^T\text{d}t\, e^{im\Omega t}t_x^{(\text{SO})}(t)=t_x^{(\text{SO})}\mathcal{J}_m(\mathcal{A}_x)(-i)^m, \label{eq: effective t-x} \\
t_y^{(\text{SO})m}=&\frac{1}{T}\int_0^T\text{d}t\, e^{im\Omega t}t_y^{(\text{SO})}(t)=t_y^{(\text{SO})}\mathcal{J}_m(\mathcal{A}_y).
\label{eq: effective t-y}
\end{align}
Here, $\mathcal{J}_m(x)$ is the spherical Bessel function of the first kind.
When $m=1$, for which the denominator on the right-hand side of Eq.~\eqref{eq: spin-dependent effective tunneling amplitude} is minimum, Eqs.~\eqref{eq: effective t-x} and \eqref{eq: effective t-y} show that the effective tunneling amplitude is proportional to $-i=e^{-i\pi/2}$ and $1$ in the $x$ and $y$ directions, respectively.
As a result, a $\pi/2$ phase emerges in the effective tunneling of electrons around a triangular closed loop made by the diagonal and two sides of the square unit cell. 
Because the time-reversed process would be associated with a $-\pi/2$ phase, the time-reversal symmetry is effectively broken.
If larger values of $m$ are used in Eq.~\eqref{eq: spin-dependent effective tunneling amplitude}, the emergent phase would be different from $\pi/2$, but the time-reversal symmetry breaking would remain unchanged, as justified by the outcome of the spin polarization computation.
As shown in Fig.~\ref{fig: spin polarization as a function of energy}a, $P_\text{s}$ changes its sign when the circular polarization of the laser field is reversed.
Moreover, $P_\text{s}$ can approach unity if $\bar{\mu}$ is within an appropriate range of values, which is narrow compared to the ordinary CISS in chiral molecules~\cite{Guo12, Guo14} because of the high frequency of the laser.
Future works will examine how the width of this energy range for $\bar{\mu}$ changes in achiral systems other than the 2D square lattice.

Chiral molecules can be combined with light-matter interactions to obtain a wider energy range for large spin polarization.
As an example, CISS in a helical molecule driven by a circularly polarized laser field is considered as follows.
A one-dimensional Hubbard model with a spatially dependent SOC-induced tunneling amplitude is used to describe the transport of electrons through a helical molecule~\cite{Guo14}.
The Hamiltonian is given by
\begin{align}
\hat{H}_\text{SO}=\sum_{j=1}^{N-1}\sum_{k=1}^{N-j}\sum_{s,s'}
\left(2i t^\text{SO}_k\cos\varphi_{j,k}^- \hat{c}^\dagger_{j+k,s}\sigma^{j,k}_{ss'}\hat{c}_{j,s'} +\text{h.c.}\right),
\end{align}
where $N$ is the number of sites, $t^\text{SO}_k$ represents the tunneling amplitude over $k$ sites, and $\sigma^{j,k}=\left(\sigma_x\sin\varphi_{j,k}^+-\sigma_y\cos\varphi_{j,k}^+\right)\sin\theta_k+\sigma_z\cos\theta_k$ with $\varphi_{j,k}^\pm=(\varphi_{j+k}\pm\varphi_j)/2$ and $\varphi_j=j\Delta\varphi$. 
Here, $\Delta\varphi$ denotes the twist angle between the nearest neighbor sites, and $\theta_k$ represents the angle between the vector connecting $k$-neighbor sites and the $x-y$ plane orthogonal to the helical axis ($z$ axis).
Long-range electron tunneling is needed for the occurrence of CISS in a single helix, indicating the significance of quantum interference between different transport pathways~\cite{Gohler11, Mishra13}.
An exponential decay of tunneling amplitude with distance is assumed.
The values of the system parameters were set to be the same as those for the $\alpha$-helical protein studied in Ref.~\cite{Guo14}.
Peierls substitution is used to describe the light-matter coupling in the same manner as above.
This guarantees that, even in the presence of a long-range tunneling, the phase acquired by an electron tunneling around a closed loop is determined by the magnetic flux piercing through the area circumvented by the loop.
The spin polarization is shown in Fig.~\ref{fig: spin polarization as a function of energy}b as a function of the average chemical potential for $N=30$.
The values of the driving frequency and dimensionless amplitude $\mathcal{A}=eA_0R/\hbar$, where $R$ is the helix radius, are the same as those in the case of a square lattice.
A wider range of energies is observed for large spin polarization than in the case of a square lattice.

\begin{figure}[tbp] 
  \centering
  \includegraphics[width=3.4in, keepaspectratio]{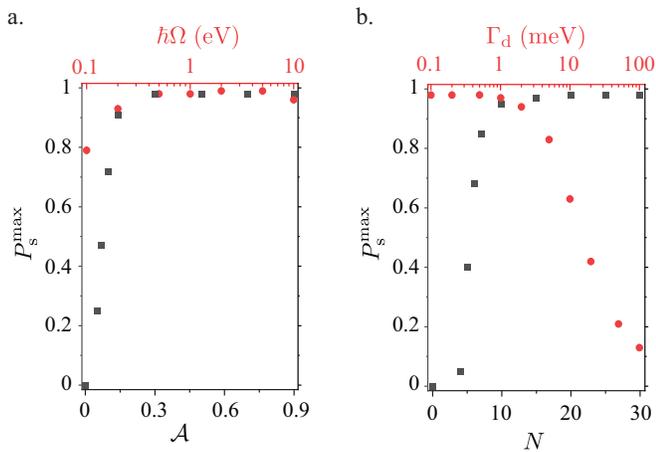}
  \caption{Dependence of maximum spin polarization $P_\text{s}^\text{max}$ of electron transport through a helical molecule driven by an external circularly polarized laser field on (a) dimensionless driving amplitude $\mathcal{A}$ and driving frequency $\Omega$ and (b) length of the molecule (the number of atomic sites $N$) and dephasing rate $\Gamma_\text{d}$.}
  \label{fig: dependence of maximum spin polarization}
\end{figure}

Figure~\ref{fig: dependence of maximum spin polarization}a shows the dependence of the maximum spin polarization (over the variable average chemical potential) on the driving amplitude and frequency. 
It is clearly seen that the spin polarization increases with the driving amplitude, and $P_\text{s}^\text{max}$ approaches unity for a dimensionless amplitude $A\gtrsim 0.3$.
In contrast, the maximum spin polarization is essentially frequency independent over a broad range $0.5\,\text{eV}\leq \hbar\Omega \leq 5\,\text{eV}$.
Figure~\ref{fig: dependence of maximum spin polarization}b shows the dependence of $P_\text{s}^\text{max}$ on the length of the molecule ($N$ atomic sites).
Evidently, the maximum spin polarization increases with $N$ and $P_\text{s}^\text{max}$ can be large, even for a relatively short molecule with $N=10$.
Additionally, when the average chemical potential of the leads deviates from the site energy of the system by an integer multiple of $\hbar\Omega$, electrons can be transported through the system driven by the laser field. 
Physically, the energy disparity is transformed into photons by light absorption or emission.

Finally, the dependence of the spin polarization on the rate of dephasing is examined. 
Dephasing is the loss of phase and spin memory of electrons due to inelastic scattering of electrons by lattice or molecular vibrations, namely phonons.
Buttiker's virtual leads are introduced at each site with broadening $\Gamma_\text{d}$ to simulate the effect of dephasing~\cite{Datta-book}.
Because there are no net currents flowing through the virtual leads, the voltages of virtual leads can be determined by using the Landauer-Buttiker formula for the current in the $q$th lead (real or virtual) with spin $s$ at energy $\epsilon$: $I_{q,s}(\epsilon)=(e^2/h)\sum_m T_{q,m,s}(\epsilon)(V_m-V_q)$, where $V_q$ denotes the voltage in the $q$th lead and $T_{q,m,s}(\epsilon)=\text{Tr}_s\left\{ \Gamma^q G^\text{R}(\epsilon)\Gamma^m G^\text{A}(\epsilon)\right\}$ is the transmission coefficient from the $m$th to the $q$th lead.
Here, $\Gamma^q$ represents the broadening matrix of the $q$th lead.
The electronic Green's functions can be obtained from the effective Hamiltonian for electrons in the high-frequency limit using $G^\text{R}(\epsilon)=[G^\text{A}(\epsilon)]^\dagger=\left[\epsilon I-\hat{H}_\text{eff}-\sum_{q}\Sigma^\text{R}_q\right]^{-1}$ with $\Sigma^\text{R}_q=-i\Gamma^q/2$.
The maximum spin polarization is shown in Fig.~\ref{fig: dependence of maximum spin polarization}b as a function of $\Gamma_\text{d}$, which is proportional to the dephasing rate.
It is clear that $P_\text{s}^\text{max}$ rises monotonically as the rate of dephasing decreases. 
This is in contrast to the ordinary CISS in chiral systems not driven by a laser field, where the spin polarization first rises with the dephasing rate before reaching its maximum value and then it starts to fall with a further increase in the dephasing rate.
As a result, it is possible to systematically increase the spin polarization, for instance, by lowering the temperature.
For the electron transport through a molecule at room temperature, where the rate of dephasing is $\Gamma_\text{d}\simeq 5\,\text{meV}$~\cite{Guo12, Guo14}, $P_\text{s}^\text{max}$ can be as high as 0.83 for $\mathcal{A}=1$.

\section{Conclusion}
\label{sec: Conclusion}
We have shown that achiral molecules and materials can be used to realize CISS in two-terminal electron transport provided that the system is driven by an external circularly polarized laser field and the average chemical potential of the two leads is within the proper range of values, which is narrow because of the high frequency of the laser.
Because the light-matter interaction can effectively break the time-reversal symmetry in the dynamics of electrons, spin polarization is nonzero without dephasing, in contrast to ordinary CISS in chiral molecules.
Chiral molecules can be combined with light-matter interactions to obtain a wider range of energies for large spin polarization. 
For sufficiently strong driving and a slow dephasing  rate, which are achievable in current experiments, the spin polarization can get close to unity.
With potential applications in spintronics and chemical reactions, Floquet-engineered CISS provides a potent tool for controlling spin dynamics in a variety of molecules and materials.


\begin{acknowledgements}
N. T. P. thanks Pham Quang Trung for fruitful discussion on numerical calculations. 
The computations were performed at the Research Center for Computational Science, Okazaki, Japan.
\end{acknowledgements}


\end{document}